# Mechanism and Kinetics of Na$^+$ Ion Depletion under the Anode during Electro-thermal Poling of a Bioactive Glass

C.R. Mariappan, B. Roling*

*Fachbereich Chemie, Physikalische Chemie, Philipps-Universität Marburg,*
*Hans-Meerwein-Straße, 35032 Marburg, Germany.*

___

**Abstract**

Electro-thermal poling experiments were carried out on a 46.4 SiO$_2$ – 25.2 Na$_2$O – 25.2 CaO – 3.2 P$_2$O$_5$ (46S4) bioactive glass, and the kinetics of the Na$^+$ ion depletion layer formation under the anode was studied in-situ by means of ac impedance spectroscopy. One important finding is a linear relation between the depletion layer thickness and the applied voltage, which is in contrast to the predictions of standard space charge theory. The average electric field in the layer is independent of the voltage and is close to the dielectric breakdown field of alkali ion conducting glasses. Furthermore, we observe that the thickness of the depletion layer is established on a much shorter time scale than the resistance. We explain these results by assuming that the huge electric fields created under the anode during Na$^+$ ion depletion lead to a strong increase of the electronic mobility in the layer and to charge compensation via extraction of electrons. It is shown that in the initial stages of the depletion process, a relative Na$^+$ ion depletion of only 400 ppm is sufficient to generate electric fields of the order of the dielectric breakdown field.

___

## 1. Introduction

In recent years, electro-thermal poling experiments on alkali ion conducting silicate and phosphosilicate glasses have been carried out with the aim to control physical, chemical and biological properties, such as second-order nonlinear optical susceptibility and bioactivity [1-6]. During a poling experiment, a glass sample is sandwiched between ion-blocking metal electrodes and is heated to a temperature $T_p$ typically between 200 and 500 °C, while applying a voltage $V_p$ in the range from a few V to a few kV. Then the sample is cooled down to room temperature under the applied voltage. At the poling temperature $T_p$, mobile alkali ions move towards the cathode (negative electrode) leaving behind a negatively charged interfacial layer under the anode (positive electrode) [3,4]. The negative charge density in this interfacial layer causes high electric fields, which lead to an effective second-order nonlinear susceptibility of the poled glasses [4]. Furthermore, it has been found that the bioactivity of poled glasses, in particular the bone binding capability, is influenced by the electrical fields and potentials in the interfacial layers under the electrodes [6,7].

\* – Corresponding author: B. Roling, Phone: +49 6421 28 22310, Fax: +49 6421 28 22309, e-mail: roling@staff.uni-marburg.de

Up to now, the electrical and electrochemical processes during the poling experiments are not very well understood. Experimentally determined values for the thickness of the alkali ion depletion layer under the anode are much larger than predicted by standard space charge theory. When, for instance, a 46.4 $SiO_2$ – 25.2 $Na_2O$ – 25.2 $CaO$ – 3.2 $P_2O_5$ (46S4) bioglass with a number density of mobile $Na^+$ ions, $N_{V,Na^+} = 1.23 \cdot 10^{28}$ m$^{-3}$, is poled with a voltage of $V_p$ = 500 V, profiling of the Na concentration by means of SEM/EDX reveals a depletion layer thickness of about 1.5 μm, while space charge theory predicts about 10 nm [8].

In this paper, we present the results of detailed ac impedance spectroscopic studies during poling experiment. These results provide strong indication that the discrepancies between standard space charge theory and experimental observation are related to field-induced electronic mobilities in the interfacial layer under the anode. In addition, the in-situ ac impedance spectra provide valuable new information about the kinetics of layer formation.

## 2. Experimental

Samples of bioactive 46S4 glass with chemical composition 46.4 $SiO_2$ – 25.2 $Na_2O$ – 25.2 $CaO$ – 3.2 $P_2O_5$ (mol%) were prepared, cut and polished as described in Ref [8]. Pt electrodes were sputtered onto both faces of the samples in order to ensure good electrical contact. AC impedance spectroscopy was carried out in a frequency range from 0.1 Hz to 1 MHz and at different temperatures using a Novocontrol Alpha-AK impedance analyzer. The analyzer is equipped with a broadband high voltage amplifier providing a maximum voltage of 2000 V (dc bias voltage + ac voltage amplitude). The sample temperature was controlled by the Novocontrol Quatro Cryosystem.

## 3. Results

### 3.1 Ac impedance spectra during poling

Figure 1 shows Nyquist plots of the ac impedance, $Z'(\nu) - i \cdot Z''(\nu)$, for a 46S4 bioglass sample at different temperatures. The spectra were recorded while heating the sample with a constant rate of 0.75 K/min. The temperature increase during the recording of a single spectrum (taking about 1 min) is small, so that the spectra are quasi isothermal. The applied dc and rms ac voltages were $V_p$ = 250 V and $V_{ac}$ = 0.5 V, respectively. At low temperatures, the Nyquist plots are characterised by a single semicircle, reflecting the bulk electrical properties of the glass sample. At temperatures above 413 K, a second low-frequency semicircle appears due to the formation of interfacial layers under the electrodes.

In order to find out whether the second semicircle is caused by the interfacial layers under both electrodes or by a single interfacial layer under one of the electrodes, we took two poled samples and removed either the interfacial layer under the anode or under the cathode by means of mechanical grinding. Then we sputtered a fresh metal electrode film on the ground face of the sample and took a new ac impedance spectrum. The results clearly revealed that the second semicircle is caused by the alkali ion depletion layer under the anode.

### 3.2 AC conductivity and permittivity spectra during poling

Interesting information about the kinetics of the interfacial layer formation under the anode can be obtained by analysing ac conductivity $\sigma'(\nu)$ and permittivity $\varepsilon'(\nu)$ spectra during poling. These spectra can be calculated from the impedance spectra by using the following relations [9]:

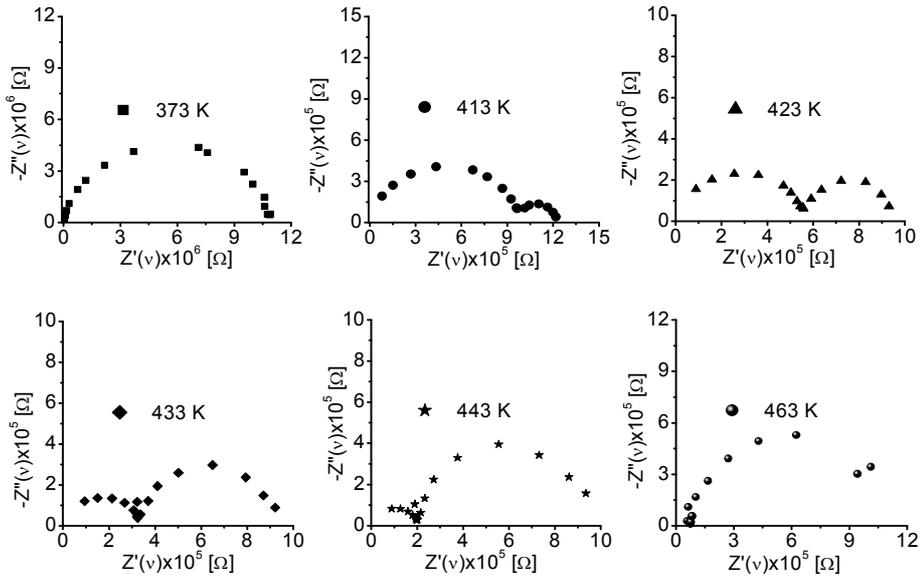

Fig.1. Nyquist plot of the ac impedance spectra of 46S4 glass during electro-thermal poling. The applied bias dc voltage is 250 V, and the rms ac voltage is 0.5 V.

$$\sigma'(\nu) = \frac{Z'(\nu)}{[Z'(\nu)]^2 + [Z''(\nu)]^2} \cdot \frac{L}{A} \quad (1)$$

$$\varepsilon'(\nu) = \frac{1}{2\pi \cdot \nu \cdot \varepsilon_0} \cdot \frac{Z''(\nu)}{[Z'(\nu)]^2 + [Z''(\nu)]^2} \cdot \frac{L}{A} \quad (2)$$

Here, $L$ and $A$ denote the sample thickness and area, respectively, while $\varepsilon_0$ denotes the permittivity of free space.

In Fig. 2 we show results for a glass sample poled with $V_p = 25$ V under a constant heating rate of 0.75 K/min. The applied rms ac voltage was $V_{ac} = 0.5$ V. Values for $\sigma'$ and $\varepsilon'$ are plotted versus temperature T at different frequencies $\nu$. At low temperatures, the isochronal $\varepsilon'$ plots are determined by the bulk permittivity of the glass. Below room temperature, $\varepsilon'$ is identical to $\varepsilon_\infty$, i.e. the high-frequency permittivity caused by vibrational and electronic polarization. Around room temperature, there is a $\varepsilon'$ step due to subdiffusive ion movements in the bulk [10,11]. At temperatures above 400 K, we find a steep rise of $\varepsilon'$ and a leveling off into a plateau with a value of about $4 \cdot 10^4$. This permittivity plateau reflects the capacitance of the interfacial layer under the anode. Thus, within in the time window considered here extending from 1 ms to 100 ms, the interfacial layer forms at temperatures between 400 K and 500 K.

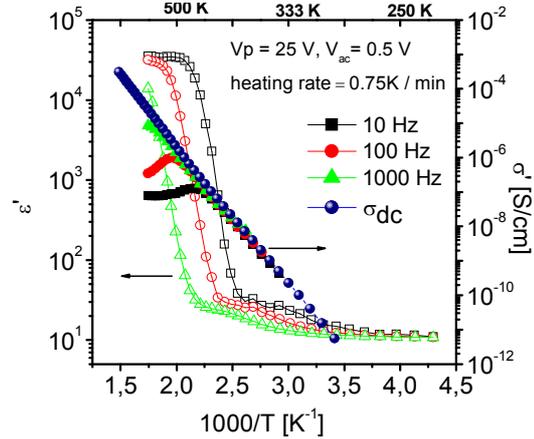

Fig. 2. Temperature-dependent permittivity and conductivity of 46S4 glass at selected frequencies during heating with a constant rate.

The conductivity $\sigma'(\nu)$ shows Arrhenius behaviour at low temperatures. At temperatures above 450 K, the conductivity drops due to the increasing resistance of the $Na^+$ ion depletion layer under the anode. Remarkably, the conductivity drop starts in a temperature range, where the permittivity has almost been reached its plateau value. With increasing temperature, the permittivity remains constant (plateau regime), while the conductivity drops further. The fact that in the isochronal plots in Fig. 2, the conductivity drop occurs at higher temperatures than the permittivity plateau formation implies that the process underlying the conductivity drop is much slower than the process underlying the permittivity plateau formation.

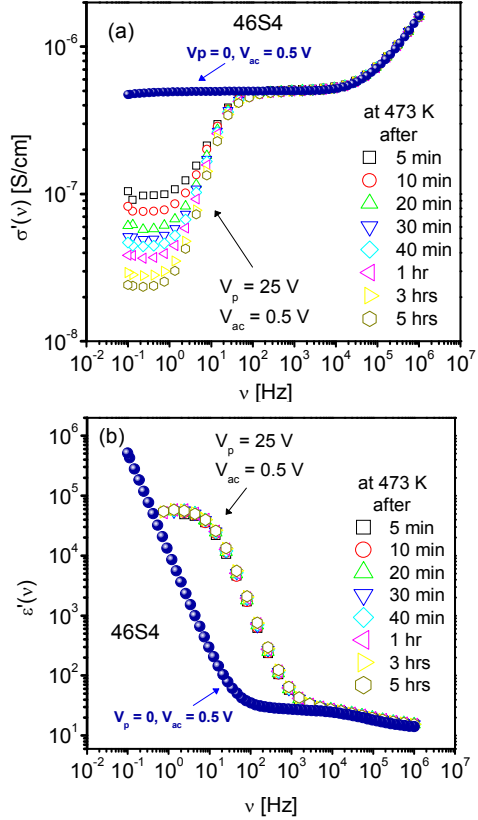

Fig. 3. (a) Ac conductivity and (b) ac permittivity spectra of 46S4 glass at various times during poling at 473 K with a dc bias voltage of 25 V. The rms ac voltage is 0.5V.

This can also be directly seen from isothermal conductivity $\sigma'(\nu)$ and permittivity $\varepsilon'(\nu)$ spectra shown in Fig. 3. The conductivity spectrum of an unpoled 46S4 glass sample is characterized by a low-frequency plateau, reflecting bulk $Na^+$ ion transport. Poling with a dc voltage of 25V leads to a low-frequency dispersion and to a low-frequency plateau reflecting the higher resistance of the interfacial layer under the anode as compared to the bulk resistance. The interfacial layer resistance increases with increasing poling time, and even after 5 h of poling the resistance is still not constant. In contrast, in the permittivity spectra, the low-frequency plateau reflecting the capacitance of the interfacial layer is constant after a much shorter time, namely after 5 min or even faster.

When we now assume that the bulk capacitance $C_{bulk}$ and the interfacial layer capacitance $C_{interface}$ are primarily determined by the sample thickness $L$ and by the interfacial layer thickness $d$, respectively, while the specific relative permittivities of bulk and interface are virtually identical, then we arrive at the following expression for the interfacial layer thickness d [12]:

$$d = \frac{L}{C_{interface} / C_{bulk}} \quad (3)$$

Thus, the capacitance of interfacial layer is a direct measure of its thickness.

### 3.3 Results for interfacial layer resistance and capacitance after poling

Figure 4 shows Arrhenius plots for the bulk dc conductivity and for the dc conductivity of interfacial layer under the anode after poling with different dc voltages. The specific conductivity of the interfacial layer was calculated by taking into account the layer thickness derived from Eq. (3). As seen from the figure, the bulk dc conductivity is Arrhenius activated with an activation energy of 0.84 eV. The interfacial layer conductivity is much lower and exhibits a much higher activation energy of 2.73 eV. Remarkably, the interfacial layer conductivity is, in a first approximation, independent of the poling voltage.

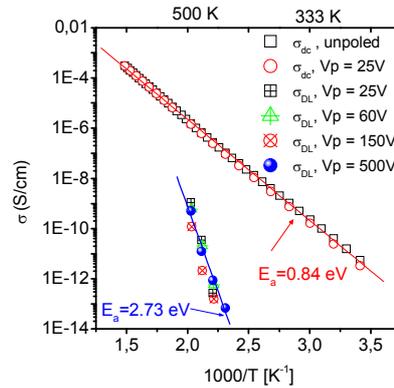

Fig. 4. Arrhenius plot of bulk dc conductivity, $\sigma_{dc}$, and of the depletion layer conductivity, $\sigma_{DL}$, for a 46S4 glass after poling with different dc voltages.

On the other hand, the thickness of the interfacial layer increases linearly with the applied voltage as shown in Fig. 5.

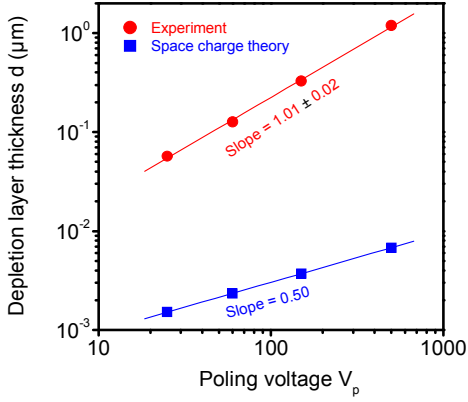

Fig. 5. Voltage dependence of the Na$^+$ ion depletion layer thickness from experiment (Eq. (3)) and predicted by standard space charge theory (Eq. (4)).

## 4. Discussion

In the bioactive 46S4 glass studied here, the Na$^+$ ions are most mobile charge carrier species. When a poling experiment is carried out, the Na$^+$ ions move away from the anode and leave behind a Na$^+$ ion depletion layer with a negative charge density. According to standard space charge theory, the thickness of the depletion layer, $d_{SCT}$ is given by [13,14]:

$$d_{SCT} = \sqrt{\frac{2 \cdot \varepsilon_0 \cdot \varepsilon \cdot V_p}{N_{V,Na^+} \cdot e}} \quad (4)$$

with $\varepsilon$ and $e$ denoting the relative bulk permittivity of the glass and the elementary charge, respectively. Eq. (4) predicts an increase of the depletion layer thickness with the square root of the poling voltage, in constrast to our experimental results revealing a linear dependence. In Fig. 5 and 6, we compare the theoretical and experimental values for the thickness and for the average electric field $E_{avg}$ in the depletion layer. The average field $E_{avg}$ was calculated by dividing the poling voltage by the layer thickness.

Space charge theory predicts fields between 10$^{10}$ V/m and 10$^{11}$ V/m, which is far above the dielectric breakdown strength of ion conducting glasses. In contrast, the experimental data for $E_{avg}$ are independent of the poling voltage and are in the range $4 - 5 \cdot 10^8$ V/m. This value is close to the dielectric breakdown strength of alkali ion conducting glasses [15, 16]. At these field strengths, field-induced electronic conduction processes are expected to take place. It seems likely that such processes are responsible for the discrepancies between standard space charge theory and experimental results.

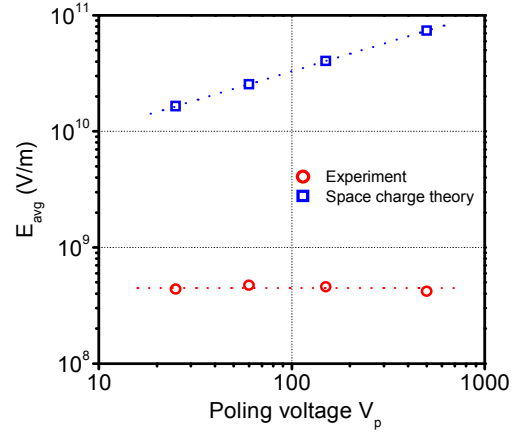

Fig. 6. Average electric field in the Na$^+$ ion depletion layer under the anode from experiment and predicted by standard space charge theory.

Therefore, we suggest the following scenario for the formation of the interfacial layer under the anode. Movements of Na$^+$ ions under the external electric field lead to an increasing negative charge density and to an increasing electric field under the anode. When the electric field approaches the breakdown field, electrons become more and more mobile. They move towards the positive electrode and are then extracted from the glass, thereby reducing the negative charge density and the electric field. Due to this charge compensation process, more Na$^+$ ions are able to move away from the anode. This process continues until the layer under the anode is completely depleted of Na$^+$ ions. It is important to note that an increase of the depletion layer thickness beyond the experimentally observed values is not possible, since this would lead to electric field strengths being too low for electronic charge compensation processes.

Our experimental results for the kinetics of the interfacial layer formation reveal different time scales of the underlying processes. While the thickness of the layer is constant after a relatively short time interval, the decrease of the layer resistance takes much longer. In order to rationalize this, we estimate in the following the time that is needed after switching on the poling voltage to reach electric fields in the range $4-5 \cdot 10^8$ V/m. According to electrostatic theory, the stored charge per area in a charged surface layer, $Q_A$, and the electric field at the surface $E_s$ are related by [17]:

$$Q_A = \varepsilon_0 \cdot \varepsilon \cdot E_S \qquad (5)$$

When the bulk dc conductivity of the glass and the externally applied electric field are denoted by $\sigma_{dc}^{bulk}$ and $E_{ext}$, respectively, the time needed for charging the layer can be written as:

$$t_{charge} = \frac{|Q_A|}{\sigma_{dc}^{bulk} \cdot E_{ext}} \qquad (6)$$

With $E_S \approx 4-5 \cdot 10^8$ V/m, $E_{ext} = 5.3 \cdot 10^4$ V/m, and $\sigma_{dc}^{bulk} = 5 \cdot 10^{-7}$ S/cm at T = 473 K, we obtain:

$$t_{charge} = \frac{\varepsilon_0 \cdot \varepsilon \cdot |E_S|}{\sigma_{dc}^{bulk} \cdot E_{ext}} \approx 15 \text{ ms} \qquad (7)$$

A temperature of 473 K was chosen for this calculation, since the impedance spectroscopy results shown in Fig. 3 were obtained at this temperature. From Eq. (7) we learn that about 15 ms after switching on the poling voltage, the electric field at the glass surface reaches values close to the breakdown strength and electronic charge compensation processes take place. Of course, a time scale of 15 ms is too short to be resolved in our time-dependent measurements of the capacitance plateau formation as shown in Fig. 3 (b). In order to obtain experimental values for $t_{charge}$, the measurements will have to be carried out at lower temperatures, where $t_{charge}$ is of the order of minutes.

Next we estimate the number of $Na^+$ ions which have to leave the interfacial layer under the anode in order to create a stored charge $Q_A = \varepsilon_0 \cdot \varepsilon \cdot E_S$ with $E_S \approx 4-5 \cdot 10^8$ V/m. With $\varepsilon \approx 10$ and $d \approx 50$ $nm$, we obtain for the average charge density $\rho$ in the layer:

$$\rho = \frac{Q_A}{d} \approx -8 \cdot 10^5 \text{ As/m}^3 \qquad (8)$$

This corresponds to a relative change in the number density of $Na^+$ ions in the depletion layer as compared to the bulk:

$$\frac{\Delta N_{V,Na^+}}{N_{V,Na^+}^{bulk}} = \frac{\rho}{N_{V,Na^+}^{bulk} \cdot e} \approx -400 \text{ ppm} \qquad (9)$$

Thus, when 400 ppm of the $Na^+$ ions leave the interfacial layer, the electric field becomes so high that electronic charge compensation processes start to take place.

On the other hand, for a complete $Na^+$ ion depletion, a 2500 times higher ionic charge flow is necessary, and accordingly, the increase of the layer resistance continues for much longer time scales. This explains why we find significantly different time scales in the capacitance and resistance spectra of the bioglass. However, for a more quantitative understanding, more detailed kinetic experiments and modeling is needed. In our opinion, such a quantitative understanding is important for achieving a quantitative control over the electric fields and potentials in the interfacial layers during electro-thermal poling experiments.

## 5. Conclusions

Ac impedance studies on 46S4 bioactive glass during electro-thermal poling yield valuable information about the mechanism and the kinetics of the $Na^+$ ion depletion layer formation under the anode. The linear dependence of the layer thickness on the poling voltage implies that the average electric field strength in the layer is independent of the voltage, the value of the field strength being close to the dielectric breakdown field of ion conducting glasses. We show that a relative $Na^+$ ion depletion of only 400 ppm as compared to the bulk is sufficient to generate internal electric fields of the order of the breakdown field. At these high field strengths, electrons are expected to become mobile, so that compensation of the negative charge density in the layer via extraction of electrons can take place. This charge compensation reduces the internal field and enables further Na$^+$ ion depletion of the layer.

When we study the capacitance and the resistance of the interfacial layer during poling via in-situ impedance spectroscopy, we find that the capacitance, and thus the thickness of the layer, is established on a much shorter time scale than the resistance. We suggest that the thickness is established on a time scale that is needed to reach electric fields of the order of the breakdown field and to induce electronic charge compensation processes. This time scale is relatively short (of the order of 10 ms), since a relative Na$^+$ ion depletion of only 400 ppm is necessary. Subsequently, the resistance of the layer drops further until a more or less complete Na$^+$ ion depletion is achieved.

**Acknowledgement**

We would like to thank the German Science Foundation (Grant Ro 1213/6-1) and the Alexander von Humboldt foundation (Research fellowship for C. R. M.) for financial support of this work.